\documentstyle [aps,preprint]{revtex}
\tightenlines
\begin{document}
\def\fnote#1#2{
\begingroup\def\thefootnote{#1}\footnote{#2}\addtocounter{footnote}{-1}
\endgroup}
\def\dslash{\not{\hbox{\kern-2pt $\partial$}}}
\def\eslash{\not{\hbox{\kern-2pt $\epsilon$}}}
\def\Dslash{\not{\hbox{\kern-4pt $D$}}}
\def\Aslash{\not{\hbox{\kern-4pt $A$}}}
\def\Qslash{\not{\hbox{\kern-4pt $Q$}}}
\def\Wslash{\not{\hbox{\kern-4pt $W$}}}
\def\pslash{\not{\hbox{\kern-2.3pt $p$}}}
\def\kslash{\not{\hbox{\kern-2.3pt $k$}}}
\def\qslash{\not{\hbox{\kern-2.3pt $q$}}}
\def\np#1{{\sl Nucl.~Phys.~\bf B#1}}
\def\pl#1{{\sl Phys.~Lett.~\bf B#1}}
\def\pr#1{{\sl Phys.~Rev.~\bf D#1}}
\def\prl#1{{\sl Phys.~Rev.~Lett.~\bf #1}}
\def\cpc#1{{\sl Comp.~Phys.~Comm.~\bf #1}}
\def\cmp#1{{\sl Commun.~Math.~Phys.~\bf #1}}
\def\anp#1{{\sl Ann.~Phys.~(NY) \bf #1}}
\def\etal{{\em et al.}}
\def\half{{\textstyle{1\over2}}}
\def\be{\begin{equation}}
\def\ee{\end{equation}}
\def\ba{\begin{array}}
\def\ea{\end{array}}
\def\tr{{\rm tr}}
\def\Tr{{\rm Tr}}
\title{Canonical quantization of a particle near a black hole
\thanks{Research
supported by the DoE under grant DE--FG05--91ER40627.}}
\author{
George Siopsis
\fnote{\ddagger}{E-mail: \tt gsiopsis@utk.edu}}
\address{Department of Physics and Astronomy, \\
The University of Tennessee, Knoxville, TN 37996--1200.\\
}
\date{May 2000}
\preprint{UTHET--00--0501}
\maketitle
\begin{abstract}
We discuss the quantization of a particle near an extreme Reissner-Nordstr\"om black hole in the canonical formalism. This model appears to be described by
a Hamiltonian with no well-defined ground state. This problem can be circumvented
by a redefinition of the Hamiltonian due to de Alfaro, Fubini and Furlan (DFF).
We show that the
Hamiltonian with no ground state corresponds to a gauge in which there
is an obstruction at the boundary of spacetime requiring a modification of the
quantization rules. The redefinition of the Hamiltonian {\em \`a la} DFF
corresponds to a different choice of gauge. The latter is a good gauge leading
to standard quantization rules. Thus, the DFF trick is a consequence of a
standard gauge-fixing procedure in the case of black hole scattering.
\end{abstract}
\renewcommand\thepage{}\newpage\pagenumbering{arabic}

\section{Introduction}

The simplest quantum mechanical system with conformal symmetry is described by the Hamiltonian
\be
H = {p^2\over 2} + {g\over 2x^2}
\ee
It is easy to see that this Hamiltonian possesses a continuous spectrum down to
zero energy and there is no well-defined ground state. A solution to this problem
was suggested by de Alfaro, Fubini and Furlan (DFF) a long time ago~\cite{bib1}. They proposed
the redefinition of the Hamiltonian by the addition of a harmonic oscillator
potential which is also the generator of special conformal transformations,
\be
K = {x^2\over 2}
\ee
The new Hamiltonian is defined by
\be H' = {1\over\omega} (H+\omega^2 K)\ee
where we introduced a scale parameter $\omega$ (infrared cutoff). $H'$ has a well-defined vacuum and a discrete spectrum, which can
actually be computed exactly,
\be
E_n = 2n+1+\sqrt{g+1/4}
\ee
Notice that the spectrum is independent of the arbitrary scale parameter $\omega$.
The supersymmetric case can be dealt with in the same way.

This problem is also encountered in the quantization of a (super)particle moving in the
vicinity of the horizon of a black hole~\cite{bib2,bib2a,bib2b,bib2c}. The near-horizon geometry of a
Reissner-Nordstr\"om black hole is $AdS_2\times S^n$. The isometries of the $AdS_2$
space are conformal symmetries of the particle. As a result, its motion is described by (super)conformal quantum mechanics whose non-relativistic limit
is of the form discussed above. The DFF redefinition of the Hamiltonian has a nice
interpretation in this case as a redefinition of the time coordinate. The DFF
Hamiltonian corresponds to a globally defined time coordinate whereas the
conformally invariant definition does not. Thus, the DFF trick appears plausible
on physical grounds.

This redefinition has also been applied to more general physical systems, such as
the scattering of extended objects~\cite{bib3} and the interaction of extreme black holes~\cite{bib4}. One again obtains a well-defined ground
state after the Hamiltonian is redefined following the DFF prescription. However,
the physical justification is not as lucid as in the single black hole case.

Here, we present an alternative derivation of the DFF procedure. We show that the
redefinition of the Hamiltonian amounts to a different choice of gauge. In the
conformally invariant case, we identify an obstruction to the standard gauge-fixing
procedure that leads to a modification of
the usual quantization rules. This obstruction comes
from the boundary of spacetime and is rooted in the fact that the
time coordinate is not defined at the boundary. On the other hand, there is no
obstruction in the choice of gauge leading to the DFF Hamiltonian. We conclude that
the DFF Hamiltonian corresponds to a good gauge choice, whereas the conformally
invariant Hamiltonian does not. Our discussion is based on the standard
Faddeev-Popov quantization procedure and is therefore applicable to more general
systems, such as the multiple black hole scattering~\cite{bib4}, as long as the system has an underlying gauge invariance.
Our results indicate that the DFF trick may arise from an obstruction to the
Faddeev-Popov procedure at some boundary of moduli space.

Our discussion is organized as follows. In Section~\ref{sec2}, we apply the
Faddeev-Popov procedure to a free particle moving in a fixed background of
curved spacetime. We also show how the procedure is equivalent to the commutation
rules one obtains from Dirac brackets. In Section~\ref{sec3}, we extend the
procedure by introducing an external electromagnetic field. In Section~\ref{sec4},
we specialize to the case of an extreme Reissner-Nordstr\"om black hole. We
show that the DFF trick is equivalent to a change of gauge. Finally, in
Section~\ref{sec5}, we summarize our conclusions and discuss possible future directions, such as the multi-black-hole moduli space.

\section{Neutral particle}
\label{sec2}

In this Section, we discuss the quantization of a particle moving in a fixed
spacetime background. First, we introduce the path integral for flat spacetime and
apply the Faddeev-Popov procedure to fix the gauge. We also show that this is
equivalent to the canonical quantization through commutation relations coming
from Dirac brackets. We then extend the discussion to general curved spacetime
backgrounds.

\subsection{Flat spacetime}

We start our discussion with the more familiar case of a particle of mass $m$
moving freely in flat spacetime. The action is
\be\label{action0}
S = \int d\tau \; L\quad, \quad
L = {1\over 2\eta} \dot x^\mu \dot x_\mu -\half\eta m^2
\ee
The dynamical variables are the spacetime coordinates, $x^\mu$ and we will be
working with the signature $(- \, +\,  +\, +)$.
Varying $\eta$, we obtain the constraint
\be
\eta^2 = - \dot x^\mu \dot x_\mu / m^2
\ee
The conjugate momenta are
\be
P_\mu = {\partial L\over \partial \dot x^\mu} = {1\over\eta}\; \dot x_\mu
\quad,\quad P_\eta = 0
\ee
The Hamiltonian is
\be
H = \dot x^\mu P_\mu - L = m\eta \chi\quad,\quad
\chi = {1\over 2m} P_\mu P^\mu + \half m
\ee
where we included a mass factor for dimensional reasons.
In terms of canonical variables, the action reads
\be\label{action}
S = \int d\tau \left( \dot x^\mu P_\mu - m\eta \chi \right)
\ee
Therefore, $\eta$ is a Lagrange multiplier enforcing the constraint
\be\label{constraint}
\chi\equiv {1\over 2m} P_\mu P^\mu + \half m = 0
\ee
which is the mass-shell condition. This constraint (analogous to Gauss's
Law in electrodynamics) generates reparametrizations of $\tau$, through Poisson
brackets,
\be\label{gauget}
\delta x^\mu = \{\, x^\mu \;,\; \chi \,\}_P \; \delta\tau = {1\over m}\,
P^\mu \delta\tau
\quad,\quad \delta P_\mu = \{\, P_\mu \;,\; \chi \,\}_P \; \delta\tau = 0
\ee
The orbits of these gauge transformations are straight lines,
\be
x^\mu = {P^\mu\over m}\, \tau + x_0^\mu
\ee
where $P^\mu, x_0^\mu$ are constant vectors. The family of orbits
in the same direction $P^\mu$ fills spacetime. Notice that we can obtain
all other families by coordinate transformations (rotations).

To quantize the system, consider the path integral,
\be
Z = {\cal N} \int {\cal D} x {\cal D} P {\cal D} \eta \; e^{iS}
= {\cal N} \int {\cal D} x {\cal D} P\; \delta(\chi)\; e^{i\int d\tau\,
\dot x^\mu P_\mu}
\ee
To define it, we need to fix the gauge
by imposing the gauge-fixing condition
\be\label{kons}
h(x^\mu) = \tau
\ee
which defines a hyper-surface that cuts each orbit precisely once.
Physically, this amounts to choosing $h(x^\mu)$ as the time coordinate.
Then its conjugate momentum, ${\cal H}$, is the Hamiltonian of the reduced
system.
%
Following the standard Faddeev-Popov procedure, we insert
\be\label{fpdet}
1 = \int {\cal D}\epsilon  \; \{h,\chi\}\; \delta (h - \{h,\chi \}\epsilon - \tau)
\ee
into the path integral and perform a reparametrization to obtain
\be
Z = {\cal N} \int {\cal D} x {\cal D} P\; \det \{h,\chi\}\;
\delta (h - \tau)\; \delta(\chi)\; e^{i\int d\tau\,
\dot x^\mu P_\mu}
\ee
We may integrate over the $\delta$-functions to reduce the dimension of phase space.
The reduced system will be described by coordinates $\overline x^i$ and
conjugate momenta $\overline P_i$. The Faddeev-Popov determinant is canceled by
the integration over $\delta(\chi)$. The momentum conjugate to $h$ (which is
identified with time) plays the
r\^ole of the Hamiltonian ${\cal H}$ of the reduced system. The path integral
becomes
\be
Z = {\cal N} \int {\cal D} \overline x {\cal D} \overline P\; e^{i\int d\tau\,
\dot{\overline x}^i \overline P_i -{\cal H}}
\ee
Equivalently, we may quantize the system in the operator formalism. To this end, we need to calculate Dirac brackets,
\be
\{\, A\;,\; B \,\}_D = \{\, A\;,\; B \,\}_P - \{\, A\;,\; \chi_i \,\}_P\;
\{\, \chi_i\;,\; \chi_j \,\}_P^{-1}\; \{\, \chi_j\;,\; B \,\}_P
\ee
where $i,j=1,2$, $\chi_1 = \chi$, $\chi_2 = h$, and promote them to commutators.

As an example, consider the special case $h(x^\mu) = x^0$ ({\em i.e.,}
identify $x^0$ with time). The reduced system is described by the coordinates
$\overline x^i = x^i$ and the Hamiltonian is
\be\label{hp0}
{\cal H} = -P_0 = \sqrt{P_iP^i+m^2}
\ee
The commutation relations we obtain from the Dirac brackets are
\be
[ P_i \;,\; x^j ] = -i \delta_i^{\; j}
\quad,\quad [{\cal H}\;,\; x^i ] = -i\, {P^i\over {\cal H}}
\ee
which are appropriate for ${\cal H}$ given by (\ref{hp0}).
Having understood the case of flat spacetime, we turn to the problem of a particle
moving in a fixed background of curved spacetime.


\subsection{Curved spacetime}

In curved spacetime, one may still use the same action as in flat space~(\ref{action0}) and derive the
form~(\ref{action}) and hence the constraint~(\ref{constraint}). The only dfference
is that indices are now raised and lowered with the background metric
$g_{\mu\nu}$ which is itself a
function of the coordinates $x^\mu$.
Therefore the gauge transformations~(\ref{gauget}) get modified to
\be
\delta x^\mu = {1\over m}\, P^\mu \;\delta\tau
\quad,\quad \delta P_\mu = {1\over m}\,\Gamma_{\nu\lambda\mu} P^\nu P^\lambda \;\delta\tau
\ee
where $\Gamma_{\nu\lambda\mu}$ are the Christoffel symbols,
\be
\Gamma_{\nu\lambda\mu} = \half(\partial_\lambda g_{\mu\nu} - \partial_\nu
g_{\lambda\mu} + \partial_\mu g_{\nu\lambda})
\ee
The orbits are the geodesics, which are solutions to the equations
\be
D\; {dx^\mu \over d\tau} \equiv {d^2x^\mu\over d\tau^2} + \Gamma_{\nu\lambda}^\mu
\; {dx^\nu\over d\tau} \; {dx^\lambda\over d\tau} =0
\ee
These can also be written in terms of the conjugate momenta,
\be
{dP_\mu\over d\tau} + \Gamma_{\nu\lambda\mu} P^\nu P^\lambda = 0
\ee
To fix the gauge, we need to find a family of surfaces parametrized by $\tau$
each member of which will intersect the geodesics precisely once.
Physically, this implies the choice of a good time coordinate.

\section{Charged particle}
\label{sec3}

If in addition to the curved background the particle interacts with an external
electromagnetic field, the action becomes
\be\label{chaction0}
S = \int d\tau \; L\quad, \quad
L = {1\over 2\eta} \dot x^\mu \dot x_\mu -\half\eta m^2
+ q \dot x^\mu A_\mu
\ee
Varying $\eta$, we obtain the constraint
\be
\eta^2 = - \dot x^\mu \dot x_\mu / m^2
\ee
The conjugate momenta are
\be
P_\mu = {\partial L\over \partial \dot x^\mu} = {1\over\eta}\; \dot x_\mu
+ q A_\mu \quad,\quad P_\eta = 0
\ee
The Hamiltonian is
\be
H = \dot x^\mu P_\mu - L = m\eta \chi\quad,\quad
\chi = {1\over 2m} \, \pi_\mu \pi^\mu + \half m\quad, \quad
\pi_\mu = P_\mu-qA_\mu
\ee
In the canonical formalism, the action reads
\be\label{chaction}
S = \int d\tau \left( \dot x^\mu P_\mu - m\eta \chi \right)
\ee
which is of the same form as in the non-interacting case~(\ref{action}).
Therefore, $\eta$ is a Lagrange multiplier enforcing the constraint
\be\label{constraint2}
\chi\equiv {1\over 2m} \, \pi_\mu \pi^\mu + \half m = 0
\ee
which is the mass-shell condition in the presence of an external vector potential. 

The orbits of the gauge transformations ($\tau$ reparametrizations) are the
trajectories of the equations of motion (Lorentz force law in curved spacetime)
\be
\dot x^\mu = {1\over m}\, \pi^\mu \quad,\quad \dot \pi_\mu + {1\over m} \Gamma_{\nu\lambda\mu} \pi^\nu\pi^\lambda = {q\over m} \, \pi^\nu F_{\mu\nu}
\quad,\quad F_{\mu\nu} = \partial_\mu A_\nu
- \partial_\nu A_\mu
\ee
or purely in terms of the coordinates $x^\mu$,
\be
\ddot x_\mu + \Gamma_{\nu\lambda}^\mu \dot x^\nu \dot x^\lambda = {q\over m} \, \dot x^\nu F_{\mu\nu}
\ee
The quantization of this system proceeds along the same lines as in the free
particle case.

\section{Extreme Reissner-Nordstr\"om black hole}
\label{sec4}

We are now ready to discuss the quantization of a particle moving near an extreme
Reissner-Nordstr\"om black hole~\cite{bib2,bib2a,bib2b,bib2c}. We will discuss both four and five spacetime
dimensions. The results in these two cases are similar.
The metric in five dimensions is
\be
ds^2 = - {1\over \psi^2} dt^2 + \psi d\vec x^2 \quad,\quad \psi = 1 +
{4Q^2\over \vec x^2}
\ee
and the vector potential is
\be
A_0 = {1\over\psi} \quad,\quad \vec A = 0
\ee
where the vectors live in a four-dimensional Euclidean space. The constant $Q$
is chosen to have the dimension of length and a factor of $4$ was including for
convenience.
Near the horizon, $\psi = 4Q^2/\vec x^2$. Using polar coordinates and switching
variables to $\psi$, we obtain
\be
ds^2 = -{1\over\psi^2} \left( dt^2 - Q^2\, d\psi^2\right)
+ 4Q^2 d\Omega_3^2
\ee
Defining
\be
x^\pm = t \pm Q \; \psi
\ee
the metric becomes
\be\label{metr4}
ds^2 = - {1\over \psi^2} \; dx^+ dx^- + 4Q^2 d\Omega_3^2
\ee
and
\be
\psi = {x^+ - x^-\over 2Q}
\ee
The vector potential has non-vanishing components
\be
A_+ = A_- = {1\over 2\psi}
\ee
In four dimensions, the metric is
\be
ds^2 = - {1\over \psi^2} dt^2 + \psi^2 d\vec x^2 \quad,\quad \psi = 1 +
{Q\over |\vec x|}
\ee
and the vector potential is
\be
A_t = {1\over\psi} \quad,\quad \vec A = 0
\ee
where the vectors live in a three-dimensional Euclidean space.
Near the horizon, $\psi = Q/|\vec x|$. Using polar coordinates and switching
variables to $\psi$, we obtain
\be
ds^2 = -{1\over\psi^2} \left( dt^2 - Q^2 d\psi^2\right)
+ Q^2 d\Omega_2^2
\ee
Defining
\be
x^\pm = t \pm Q \; \psi
\ee
the metric becomes
\be
ds^2 = - {1\over \psi^2} \; dx^+ dx^- + Q^2 d\Omega_2^2
\ee
This is of the same form as in five spacetime dimensions (Eq.~(\ref{metr4})),
apart from the scale factor in the spherical part of the metric. Thus in both
four and five dimensions,
spacetime factorizes into a product $AdS_2\times S^n$ ($n=2,3$, respectively).
Henceforth, we shall work
with $AdS_2$.
The only non-vanishing connection coefficients are $\Gamma_{\pm\pm}^\pm =
\partial_\pm \ln |g_{+-}|$. Therefore,
the geodesic equations for $x^\pm$ are
\be
\ddot x^\pm \pm  (\ln |g_{+-}|)' (\dot x^\pm)^2 = \pm {q\over m} \; \dot x^\pm F_{+-}
\ee
where $A=A_+=A_-$, and $(\ln |g_{+-}|)' = \partial_+ \ln |g_{+-}| = - \partial_- \ln |g_{+-}|$.
It is easy to see that the orbits with constant $\psi$ are geodesics,
\be
x^+ = x^-+ 2\alpha Q
\ee
where $\psi = \alpha$, parametrized by
\be
\tau = {2m\alpha\over q} \, (x^++x^-)
\ee
Motion along these geodesics is generated by the conjugate momentum,
\be H = P_++P_-\ee
%
Using $\psi\, {dA\over d\psi} = -A$, $\psi\, {dg_{+-}\over d\psi} = - 2g_{+-}$,
$F_{+-} = 2\partial_+A$,
it is straightforward to show that the following quantities are gauge-invariant
(constant along geodesics)
\be
H = P_++P_-\quad,\quad D = 2x^+P_++2x^-P_-\quad,\quad
K = (x^+)^2P_+ + (x^-)^2P_-
\ee
They obey an $SL(2,{\bf R})$ algebra
\be
\{\,H\;,\; D\,\} = -2H\quad,\quad \{\,H\;,\;K\,\} =-D\quad,\quad
\{\,K\;,\;D\,\} = 2K
\ee
reflecting the symmetry of the $AdS_2$ spacetime.
$H, D$, and $K$ generate time translations, dilatations, and special conformal
transformations, respectively.
The brackets may be Poisson or Dirac, so this is also an algebra of the gauge-fixed
system, as expected.
The constraint (generator of gauge transformations)
$\chi\equiv {1\over 2m} \pi_\mu \pi^\mu + {1\over 2} m = 0$ reads
\be
-4 \psi^2 P_+P_- +2q\psi (P_++P_-) +{L^2\over Q^2} -q^2+m^2 =0
\ee
where $L^2 = \hat g^{ij} P_iP_j$ is the square of the angular momentum operator.
The simplest gauge-fixing condition to impose is
\be\label{gauge1}
h(x^+,x^-) = x^++x^- =\tau
\ee
In this case, the Hamiltonian is
\be
{\cal H} = -H = -P_+-P_-
\ee
Using the constraint, we obtain
\be\label{eq54}
{\cal H} = {1\over \psi} \, \left(-q+\sqrt{m^2+ (\psi^2P_\psi^2+L^2)/Q^2}\right)
\ee
The other two operators in the $SL(2,{\bf R})$ algebra can be written as
\be
D = 2\hat tH + 2\psi P_\psi \quad,\quad K = (Q^2\psi^2 + \widehat{t^2}) H + 2\hat t\psi P_\psi
\ee
where $\hat t$ and $\widehat{t^2}$ are operators satisfying $\dot{\hat t} \equiv
\{ \hat t\,,\, H\} = 1$ and $\dot{\widehat{t^2}} \equiv \{ \widehat{t^2}\,,\, H\} =
2\hat t$, respectively. After some algebra,
we find
\be
\hat t = \psi P_\psi /H \quad,\quad \widehat{t^2} = Q^2\psi^2 -2qQ^2 \psi /H -
2\psi^2 P_\psi^2 /H^2
\ee
Therefore,
\be
D = 4\psi P_\psi \quad,\quad K = 2Q^2\psi^2 H -2qQ^2\psi
\ee
In the non-relativistic limit,
\be
{\cal H} = {\psi P_\psi^2\over 2mQ^2} + {g\over 8Q^2\psi}\quad,\quad
D = 4\psi P_\psi \quad,\quad K= (g/4 - 2qQ^2)\psi
\ee
where $g = 8Q^2(m-q)+4L^2/m$. Let us introduce a new variable $x$,
defined by
\be
\psi = {x^2\over 4Q^2}
\ee
In the limit $q\to m$, $Q\to \infty$, so that $Q^2(m-q)$ remains fixed,
we have
\be\label{eq58}
{\cal H} = {P^2\over 2m} + {g\over 2x^2}\quad,\quad
D = 2x P \quad,\quad K= -\half m x^2
\ee
where $P$ is the momentum conjugate to $x$.
This system does not have a well-defined vacuum.
The question then arises whether the underlying theory is inherently sick.
One may apply the DFF trick to produce a Hamiltonian system with a well-defined
ground state. The DFF trick can be understood in this case as a different choic of
time coordinate leading to a different Hamiltonian.
From our point of view, any two choices of time coordinates should be equivalent to
each other, for they merely correspond to different gauge choices.
Since the underlying theory is gauge-invariant, all gauge choices should be
equivalent to each other.

Before we discuss the vacuum problem in conjunction with the gauge-fixing procedure,
we shall introduce a class of gauges that lead to a Hamiltonian system with a
well-defined ground state.
Let the gauge-fixing condition be
\be\label{gftan}
h(x^+,x^-) = \arctan \left( {\omega x^++\omega x^-
\over 1-\omega^2 x^+x^-} \right) = \tau
\ee
where $\omega$ is an arbitrary scale. Differentiating with respect to $\tau$,
we obtain
\be
\partial_+ h \, \dot x^+ + \partial_- h \, \dot x^- =1\quad,\quad
\partial_\pm h = {\omega\over 1+\omega^2 (x^\pm)^2}
\ee
The Hamiltonian is
\be
{\cal H} = - {P_+\over\partial_+ h} - {P_-\over\partial_- h} = -{1\over\omega}
\, (H + \omega^2 K)
\ee
In the non-relativistic limit,
\be
{\cal H} = {P^2\over 2m\omega} + {g^2\over 2\omega x^2} + \half m
\omega x^2
\ee
which has a well-defined vacuum. All these gauges are of course equivalent.
Therefore, no physical quantities should depend on the scale parameter $\omega$.
In particular, notice that the spectrum in the non-relativistic limit is
independent of $\omega$.

It is puzzling that there exists a gauge (Eq.~(\ref{gauge1})) in which the vacuum is not well-defined.
To resolve the puzzle, let us follow the Faddeev-Popov gauge-fixing procedure a
little more carefully. We need to insert~(\ref{fpdet}) into the path integral
and then perform an inverse gauge transformation to eliminate the gauge parameter.
In doing so, we encounter an obstruction at the boundary of spacetime.
Under a gauge transformation, the change in the action is
\be
\delta S = \int d\tau \; {d\over d\tau} (\delta x^\mu P_\mu) - \int d\tau \,
\epsilon \dot\chi
\ee
Since $\dot\chi = \{\chi \,,\, \chi\} =0$, we conclude that
the action changes by a total derivative,
\be
\delta S = \int d\tau \; {d\over d\tau} (\delta x^\mu P_\mu)
\ee
We have
\be
\delta x^\mu = \{ x^\mu\,,\, \chi\} \, \epsilon = {\partial\chi\over \partial P_\mu} \, \epsilon
\ee
therefore,
\be
\delta S = P_\mu {\partial\chi\over \partial P_\mu} \, \epsilon \, \Bigg|_\partial
\ee
Notice that, if the generator of gauge transformations, $\chi$,
is quadratic in the momenta (as in the free particle
case~(\ref{constraint})), then the boundary contribution vanishes after imposing the constraint
$\chi =0$. In our case, $\chi$ (Eq.~(\ref{constraint2})) is not quadratic in the momenta, due to the
presence of the vector potential. Therefore $\delta S\ne 0$ and
we cannot in general get rid of the gauge parameter on the boundary
of spacetime. In the set of gauges~(\ref{gftan}), there is no boundary
contribution, because the Faddeev-Popov determinant vanishes there. Indeed,
\be
\{\, h\,,\, \chi \,\} \propto {P_+\over 1 + \omega^2 (x^+)^2} + {P_-\over 1 + \omega^2 (x^-)^2}
\ee
which vanishes as $x^\pm\to\infty$. For the gauge~(\ref{gauge1}), we obtain
a boundary contribution to the path integral,
\be
\int_\partial d\epsilon d^Dx d^DP \, \{\, h\,,\, \chi \,\}
\, \delta (h-\tau) \delta(\chi)\, \exp\left\{ iP_\mu {\partial\chi\over \partial P_\mu} \epsilon \right\}
\ee 
This obstruction is absent when $\{\, h\,,\, \chi \,\} =0$ at the boundary.
Physically, this condition implies that the boundary of spacetime is invariant under transformations
generated by $h$, which is the time coordinate after gauge-fixing ($h=\tau$).
In other words, the boundary is fixed under time translations. Thus the
time coordinate~(\ref{gauge1}) is not a good global coordinate and leads to an
obstruction in the gauge invariance of the theory. Integrating over the gauge
parameter, we obtain an additional constraint at the boundary,
\be
P_\mu {\partial\chi\over \partial P_\mu}\; \Bigg|_\partial = 0
\ee
This alters the standard commutation relations and the eigenvalue problem for
the Hamiltonian (\ref{eq54}) or (\ref{eq58}). We have not carried out an explicit
computation. This would involve the introduction of a regulator which would break
gauge invariance. Nevertheless, the resulting system should be equivalent to the
one obtained through the other choices of gauge due to the gauge invariance of the
theory.

To summarize, the na\"\i ve identification of the time coordinate~(\ref{gauge1})
leads to an obstruction in the gauge-fixing procedure for the path integral.
If this obstruction is accounted for by an appropriate modification of the
commutation relations, this choice of the time coordinate leads to a well-defined
Hamiltonian problem. The Hamiltonian system thus obtained is equivalent to
applying the DFF trick~\cite{bib1}, or identifying the time coordinate as in Eq.~(\ref{gftan})~\cite{bib2c}.
The latter is merely a different gauge choice in a gauge-invariant theory.

\section{Conclusions}
\label{sec5}

We considered the problem of quantization of a charged particle in the vicinity
of the horizon of an extreme Reissner-Nordstr\"om black hole. In this case, the
vacuum is not well-defined unless a special choice of the time coordinate is
made~\cite{bib2c}. This has been shown to be equivalent to the DFF trick for conformal
quantum mechanics~\cite{bib1}. We approached this problem through the path integral and
the Faddeev-Popov gauge-fixing procedure. We showed that the DFF trick can be
understood in terms of the standard Faddeev-Popov procedure.

We started with a general discussion of the quantization of a particle in the presence of a background
metric field as well as an external vector potential. We performed the standard
Faddeev-Popov procedure in the canonical formalism and showed its connection
to commutation relations through Dirac brackets. We then applied the procedure to
the case of interest (extreme Reissner-Nordstr\"om black hole). We showed that
the na\"\i ve identification of time coordinate (which leads to a Hamiltonian
system with no well-defined ground state) corresponds to a gauge which is not
``good." We found that in this gauge the Faddeev-Popov procedure encounters an
obstruction at the boundary of spacetime introducing an additional constraint
there. This alters the standard commutation relations and the eigenvalue problem
for the attendant Hamiltonian system. We did not calculate the effects of this
obstruction explicitly. This would require the introduction of a regulator which
would break gauge invariance explicitly and therefore alter the commutation
rules. Instead, we exhibited another set of gauges where no obstruction existed
on the boundary. We showed that this set of gauges led to a Hamiltonian system
with a well-defined vacuum, equivalent to the one obtained through the DFF trick~\cite{bib1}.

A choice of gauge is equivalent to a choice of time coordinate, because gauge
transformations are reparametrizations of the proper time of the particle.
The above discussion shows that all choices of the time coordinate are
equivalent since they correspond to different gauge choices in a gauge-invariant
theory. It would be interesting to apply our procedure to multiple black hole
scattering, where the study of moduli space seems to necessitate the introduction
of the DFF trick~\cite{bib4}. It is hard to see how this can be justified in terms of a
``proper" choice of a time coordinate (which time coordinate?). However, the underlying theory is a gauge
theory, so the Faddeev-Popov procedure should be applicable. Our results indicate
that one may encounter an obstruction to the implementation of the procedure,
perhaps on some boundary of moduli space, which would lead to an alteration
of the na\"\i ve Hamiltonian system similar to the DFF trick. Work in this
direction is in progress.

\end{document}